\documentclass[prl,tighten,article,twocolumn,showpacs,amssymb,amsmath,latexsym]{revtex4}

\newtheorem{thm}{Theorem}
\newtheorem{lem}{Lemma}

\newtheorem{proposition}{Proposition}

\newcommand{\defeq}{\ensuremath{\stackrel{\rm def}=}} 

\renewcommand{\H}{\ensuremath{{\cal H}}}

\newcommand{\X}{\ensuremath{{\cal X}}}

\newcommand{\Hn}{\ensuremath{{\cal H}^{\otimes n}}}
\newcommand{\rhon}{\ensuremath{\rho^{\otimes n}}}
\newcommand{\sigman}{\ensuremath{\sigma^{\otimes n}}}

\newcommand{\Tr}{\ensuremath{\mbox{\rm Tr}}}

\newcommand{\argmax}{\ensuremath{\mathop{\rm arg \; max}}}

\newcommand{\oneover}[1]{\ensuremath{\frac{1}{#1}}}

\begin{document}
\title{Error Exponent in Asymmetric Quantum Hypothesis Testing
and\\ Its Application to Classical-Quantum Channel coding}
\author{Masahito Hayashi}
\affiliation{ERATO-SORST Quantum Computation and Information Project,
Japan Science and Technology Agency,
201 Daini Hongo White Bldg.
5-28-3, Hongo, Bunkyo-ku, Tokyo 113-0033, Japan.
\email{masahito@qci.jst.go.jp}\\
Superrobust Computation Project,
Information Science and Technology Strategic Core (21st Century COE by 
MEXT),
Graduate School of Information Science and Technology,
The University of Tokyo,
7-3-1, Hongo, Bunkyo-ku, Tokyo, 113-0033, Japan.}


\begin{abstract}
In the simple quantum hypothesis testing problem,
upper bound with asymmetric setting
is shown by using a quite useful inequality by 
Audenaert et al\cite{ACMMABV}, quant-ph/0610027,
which was originally invented for symmetric setting.
Using this upper bound, 
we obtain the Hoeffding bound,
which are identical with the classical counter part
if the hypotheses,
composed of two density operators, are mutually commutative.
Our upper bound improves the bound by Ogawa-Hayashi\cite{Oga-Hay},
and also provides
a simpler proof of the direct part of the quantum Stein's lemma.
Further, using this bound, we obtain 
a better exponential upper bound of the average error probability of 
classical-quantum channel coding.
\end{abstract}
\maketitle


\section{Introduction}
\label{sec:intro}
One of the main difficulties appearing in quantum information theory
lies in the non-commutativity.
Hence, for further development of quantum information theory,
it is needed to accumulate the methods to resolve such difficulties.
Simple quantum hypothesis testing is 
the simplest problem describing this kind of difficulty\cite{Hiai-Petz,Ogawa-Nagaoka-2000}
because this problem is discriminating two quantum states 
(the null hypothesis and the alternative hypothesis)
as the candidates of the true state.
This problem is also the fundamental tool 
for other problems in quantum information theory.
For example, classical-quantum channel coding 
\cite{H-N,Ogawa-Nagaoka-ISIT2002},
classical-quantum wire-tap channel coding \cite{H-info},
and 
quantum fixed-length source coding (Schumacher coding\cite{Schumacher})
\cite{Hayashi-fixed-length-source,Nag-Hay}
can be analyzed through simple quantum hypothesis testing
In the single-copy case, 
simple quantum hypothesis testing 
has been solved by using quantum Neyman-Pearson Lemma 
in Holevo\cite{Ho72} and Helstrom \cite{Hel}.
However, when the number $n$ of samples is large,
the asymptotic behavior of the performance of this problem 
has been partially solved.
Several problems have been still open.

In the asymptotic framework,
Chernoff bound\cite{Cher}, 
Stein's lemma, 
Hoeffding bound\cite{Hoeffding}, Han-Kobayashi bound\cite{HK} 
are known as the bounds of the classical simple hypothesis testing.
We usually focus on the two kinds of error probabilities,
i.e., the first kind of error probability 
(the null hypothesis is rejected despite of being correct)
and the second kind of error probability 
(the alternative hypothesis is rejected despite of being correct).
Chernoff bound gives the optimal decreasing rate of the average of
these error probabilities in the symmetric setting.
In Stein's lemma, we focus on the optimal decreasing rate of
the second error probability under the constant constraint 
for the first error probability.
In Hoeffding bound, we treat the same optimal decreasing rate
under the exponential constraint for the first error probability.
That is, in this case, we treat the discriminating problem in the
{\it asymmetric} setting.
In fact, Hoeffding bound is more useful than Stein's Lemma 
for the approximation in the finite-sample case.
When the exponential constraint for the first error probability
is too strong,
the second error probability goes to 1.
That is, the $1$ minus the second error probability goes 0.
Han-Kobayashi bound gives the minimum decreasing rate 
of this value.
This exponent is often called the strong converse exponent.
Further, information spectrum approach is 
known as an effective method for general sequence of information sources.
The treatment of the difficulty due to non-commutativity
is necessary for the quantum extensions of these results.

Now, we trace the history of this research area.
First, the quantum extension of Stein's lemma
has been solved by Hiai-Petz\cite{Hiai-Petz} and 
Ogawa-Nagaoka\cite{Ogawa-Nagaoka-2000}.
The upper bound of the quantum extension of Han-Kobayashi bound 
has been obtained Ogawa-Nagaoka\cite{Ogawa-Nagaoka-2000}.
Their proof was extensively simplified by Nagaoka\cite{Nagaoka-converse}.
Hayashi improved their bound 
and obtained the tight strong converse exponent
in the quantum setting in Chapter 3 of \cite{H-info}.
The quantum extension of information spectrum approach 
wa obtained by Nagaoka-Hayashi\cite{Nag-Hay}.
Concerning the symmetric setting,
Hayashi obtained quantum Chernoff bound 
in Chapter 3 of \cite{H-info} 
when two hypothesis are unitarily equivalent with each other.
Nussbaum \& Szko\l{}a \cite{NS} obtained its lower bound.
Quite recently, Audenaert et al \cite{ACMMABV} showed that
the bound by Nussbaum \& Szko\l{}a \cite{NS} can be attained.
In their proof, they derived a quite useful inequality 
(Lemma \ref{lem:4} in this paper).

However, concerning the quantum extension of Hoeffding bound,
only a lower bound has been obtained by Ogawa-Hayashi\cite{Oga-Hay}.
Their approach is valid only in the finite dimensional case.
Also, their bound does not work effectively in the pure states case.
They also suggested the existence of a tighter lower bound.
Hence, tighter lower bounds of these problems has been desired.
In this paper, we obtain tighter lower bounds 
of Hoeffding bound
by using an extremely powerful inequality by Audenaert et al \cite{ACMMABV}.
This method is valid even in the infinite-dimensional case.
As a byproduct,
a simpler proof of the quantum Stein's lemma is also given.

Fortunately, such an asymmetric treatment of hypothesis testing
is closely related to classical-quantum channel coding,
i.e., the problem of transmitting classical information via quantum channel.
In this problem, the asymptotic transmitting rate is obtained by 
Holevo \cite{HoCh} and Schumacher-Westmoreland \cite{SW}.
However, there is no good upper bound of error probability 
with a good finite-length code.
Hayashi-Nagaoka \cite{H-N} 
derived a good relation between 
this problem and the asymmetric treatment of hypothesis testing.
In this paper, we apply this relation to our result and obtain 
a good error exponent of the average error probability
of classical-quantum channel coding,
and obtain a better and more natural 
exponential decreasing rate of error probability
than Hayashi-Nagaoka \cite{H-N}'s rate.

In the following, we outline briefly significant results in
classical hypothesis testing
for probability distributions $p^n(\cdot)$ versus $q^n(\cdot)$,
where $p^n(\cdot)$ and $q^n(\cdot)$ are
independently and identically distributed (i.i.d.) extensions
of some probability distributions $p(\cdot)$ and $q(\cdot)$
on a finite set $\X$.
In the classical case,
the asymptotic behaviors of the first kind error probability
$\alpha_n$ and the second kind error probability $\beta_n$
for the optimal test were studied thoroughly as follows.

First, when we focus on the average error concerning these 
two error probabilities in the symmetric setting,
it is natural to focus on Chernoff\cite{Cher}'s characterization:
\begin{align*}
\lim_{n\rightarrow\infty}\frac{-1}{n}\log\min\frac{\beta_n +\alpha_n}{2}
=
\max_{0 \le s \le 1} - \phi(s),
\end{align*}
where $\phi(s)$ is defined as
$\phi(s) \defeq \sum_{x\in\X} p(x)^{1-s} q(x)^{s}$.
Its quantum extension has been done by 
Nussbaum \& Szko\l{}a \cite{NS} and Audenaert et al \cite{ACMMABV}.
However, in order to treat the asymmetric setting,
we need another formulation.
When $\alpha_n$
satisfies the constant constraint $\alpha_n\le\epsilon$ $(\epsilon>0)$,
the error exponent of $\beta_n$ for the optimal test is written
asymptotically as
\begin{align}
\lim_{n\rightarrow\infty}\frac{-1}{n}\log
\min\{\beta_n |\alpha_n\le\epsilon\}
= D(p||q)
\label{classical-Stein}
\end{align}
for any $\epsilon$, where $D(p||q)$ is the Kullback-Leibler divergence.
The equality \eqref{classical-Stein} is called
Stein's lemma (see {\it e.g.} \cite{Blahut-text}, p.115).
When $\alpha_n$ satisfies the exponential constraint
$\alpha_n\le e^{-nr}$ $(r> 0)$, the error exponent of $\beta_n$
for the optimal test is asymptotically determined by
the Hoeffding bound \cite{Hoeffding}:
\begin{align*}
\varlimsup_{n\rightarrow\infty}\frac{-1}{n}\log\beta_n
= \max_{0 < s \le 1}\frac{-\phi(s)-(1-s)r}{s}.
\end{align*}
In this paper, we treat their quantum extension.
After discussing this topic, we proceed to its application to 
classical-quantum channel coding.


\section{Formulation and Main Results}
\label{sec:definition}

Let $\H$ be a Hilbert space which represents a physical system in interest.
We study the simple hypothesis testing problem for
the null hypothesis $H_0 : \rhon$
versus the alternative hypothesis $H_1 : \sigman$,
where $\rhon$ and $\sigman$ are the $n$th
tensor powers of arbitrarily given density operators 
$\rho$ and $\sigma$ on $\H$.

The problem is to decide which hypothesis is true
based on the data drawn from a quantum measurement,
which is described by a positive operator valued measure (POVM)
on $\Hn$, i.e.,
a resolution of identity $\sum_i M_{n,i} = I_n$
by nonnegative operators $M_n=\{M_{n,i}\}$ on $\Hn$.
If a POVM consists of projections on $\Hn$,
it is called a projection valued measure (PVM).
In the hypothesis testing problem, however,
it is sufficient to treat a two-valued POVM $\{M_0,M_1\}$,
where the subscripts $0$ and $1$ indicate the acceptance
of $H_0$ and $H_1$, respectively.
Thus, a hermitian matrix $T_n$ satisfying inequalities
$0\le T_n\le I$ is called a test
in the sequel, since $T_n$ is 
identified with the POVM $\{T_n,\, T_n^c\}$. 
For a test $T_n$, the error probabilities of the first kind and 
the second kind are, respectively, given by
$\Tr[\rhon T_n^c]$ and $\Tr[\sigman T_n]$,
where $T^c:=I-T$.

Next, we consider this problem in an asymmetric framework.
Let us define the optimal value for $\Tr[\rhon T_n^c]$
under the constant constraint on $\Tr[\sigman T_n]$:
\begin{align*}
\beta_n^*(\epsilon)\defeq\min
\bigl\{ \Tr[\rhon T_n^c &\bigm| A_n:\text{test},\,
\Tr[\sigman T_n]\le\epsilon \bigr\},
\end{align*}
and let
\begin{align*}
D(\rho\|\sigma)\defeq\Tr[\rho(\log\rho-\log\sigma)],
\end{align*}
which is called the quantum relative entropy.
Then we have the following theorem, which 
is obtained by Hiai-Petz\cite{Hiai-Petz} and Ogawa-Nagaoka\cite{Ogawa-Nagaoka-2000}.
\vspace{2ex}
\begin{proposition}[The quantum Stein's lemma]
For $0<\forall\epsilon<1$, it holds that
\begin{align}
\lim_{n\rightarrow\infty}\oneover{n}\log\beta_n^*(\epsilon)=-D(\rho\|\sigma).
\label{Stein}
\end{align}
\end{proposition}
\vspace{2ex}
This lemma can be proved by composing of two inequalities,
the direct part and the converse part.
The direct part is given by 
$B(\rho\|\sigma) \ge D(\rho\|\sigma)$,
and the converse part is given by 
$B^\dagger(\rho\|\sigma) \le D(\rho\|\sigma)$,
where
\begin{align*}
&B(\rho\|\sigma)\defeq 
\sup_{\{T_n\}}
\left\{ \left.
\varliminf_{n\rightarrow\infty}   \frac{-\log \Tr \sigma^{\otimes n} T_n}{n} 
\right| \lim_{n\rightarrow\infty}   \Tr \rho^{\otimes n} T_n^c =0
\right\}, \\
&B^\dagger(\rho\|\sigma)\defeq 
\sup_{\{T_n\}}
\left\{ \left.
\varliminf_{n\rightarrow\infty}   \frac{-\log \Tr \sigma^{\otimes n} T_n}{n} 
\right| \varliminf_{n\rightarrow\infty}  \Tr \rho^{\otimes n} T_n >0
\right\}  .
\end{align*}

For a further analysis of the direct part,
we focus on the decreasing exponent of the
error probability of the first kind under
an exponential constraint for the error probability of the second kind.
For this purpose, we define 
\begin{align*}
&B(r|\rho\|\sigma)\\
\defeq &
\sup_{\{T_n\}}
\left\{ \left.
\varliminf_{n\rightarrow\infty}   \frac{-\log \Tr \rho^{\otimes n} T_n^c}{n} 
\right|
\varliminf_{n\rightarrow\infty}   \frac{-\log \Tr \sigma^{\otimes n} T_n \ge r}{n}
\right\},
\end{align*}
where $\phi(s|\rho\|\sigma)\defeq
\log \Tr \rho^{1-s}\sigma^s$.
Then, we obtain the following theorem.

\begin{thm}\label{thm2}
The inequality 
\begin{align}
B(r|\rho\|\sigma)\ge
\sup_{0\le s \le 1}\frac{-s r- \phi(s|\rho\|\sigma)}{1-s}
\label{6-25-1}
\end{align}
holds.
\end{thm}
In fact, Ogawa-Hayashi \cite{Oga-Hay} obtained 
the following lower bound of 
$B(r|\rho\|\sigma)$:
\begin{align}
B(r|\rho\|\sigma)\ge
\max_{0\le s \le 1}\frac{-s r- \tilde{\phi}(s|\rho\|\sigma)}{1-s},
\end{align}
where
\begin{align}
\tilde{\phi}(s|\rho\|\sigma)\defeq
\Tr \rho \sigma^{s/2}\rho^{-s}\sigma^{s/2}.
\end{align}
As is shown in Section V of Ogawa-Hayashi \cite{Oga-Hay}
our lower bound 
$\max_{0\le s \le 1}\frac{-s r- \phi(s|\rho\|\sigma)}{1-s}$
is greater than
their bound
$\max_{0\le s \le 1}\frac{-s r- \tilde{\phi}(s|\rho\|\sigma)}{1-s}$.
The inequality (\ref{6-25-1}) was treated as an open problem in their 
paper.

\section{Proof of Main Theorems}
In the following we abbreviate 
$\phi(s|\rho\|\sigma)$ to $\phi(s)$. 
In order to prove Theorem \ref{thm2},
we use the following lemma.

\begin{lem}\label{l1}
For any two positive-semidefinite operators $X,Y$
and a real number $0 \le s \le 1/2$,
we define the projection $P$ as the projection on
the range of $X^{1-s} - Y^{1-s}$.
Then,
\begin{align*}
\Tr X^{s} Y^{1-s}\ge &\Tr \{X^{1-s} - Y^{1-s} \ge 0\} Y \\
&+ 
\Tr \{X^{1-s} - Y^{1-s} < 0\}X,
\end{align*}
where for any Hermite matrix $C$
we denote the projection 
$\sum_{c_i \ge 0 }E_i$ ($\sum_{c_i < 0 }E_i$ )
by 
$\{C \ge 0\}$ ($\{C < 0\}$) with the spectral decomposition
$C= \sum_i c_i E_i$.
\end{lem}
Only the case of $s=1/2$ has been proved in Chapter 3 of Hayashi \cite{H-info}.

Substituting $\rhon$ and $\sigman e^{-na}$ to 
$Y$ and $X$ in this lemma,
the projection $T_{n,s}:=\{(\sigman e^{-na})^{1-s} -(\rhon)^{1-s} < 0\}$
satisfies 
\begin{align}
&\Tr \sigman T_{n,s}
= \Tr X T_{n,s} e^{na} \nonumber\\
\le &\Tr X^s Y^{1-s}  e^{na}
= e^{n(1-s)a} e^{n \phi(s)} \label{6-25-2}\\
&\Tr \rhon (I-T_{n,s})= \Tr Y (I-T_{n,s})\nonumber\\
\le & \Tr X^s Y^{1-s}
= e^{-nsa} e^{n \phi(s)} \label{6-25-3}
\end{align}
for $0 \le s \le 1/2$.
For $1/2 \le t \le 1$,
the projection $T_{n,t}:=\{(\sigman e^{-na})^{t} -(\rhon)^{t} < 0\}$
satisfies 
\begin{align}
\Tr \sigman T_{n,t}
&\le e^{n(1-t)a} e^{n \phi(t)} \label{6-25-2-b}\\
\Tr \rhon (I-T_{n,t})
& \le e^{-nta} e^{n \phi(t)} \label{6-25-3-b},
\end{align}
where we substitute $1-t$, $\rhon$, and $\sigman e^{-na}$
into $s$, $X$, and $Y$.

Hence, we can easily prove the direct part of
quantum Stein's lemma, i.e., $B(\rho\|\sigma)\ge D(\rho\|\sigma)$ from 
Lemma \ref{l1}.
Putting $a= - D(\rho\|\sigma) + \epsilon$,
we obtain 
$-sa +\phi(s)\cong -\epsilon s < 0$ and
$(1-s)a +\phi(s)< - (D(\rho\|\sigma) -\epsilon) (1-s)$.
Hence, by choosing $s$ to be sufficiently small,
we obtain $B(\rho\|\sigma)\ge D(\rho\|\sigma)$.

We also choose $s_r \defeq 
\argmax_{0\le s \le 1}\frac{-s r- \phi(s|\rho\|\sigma)}{1-s}$.
Then, we have
\begin{align*}
r &= (s_r -1)\phi'(s_r) -\phi(s_r) \\
\max_{1\ge s'\ge 0}
\frac{-s' r- \phi(s')}{1-s'}
& = s_r \phi'(s_r) -\phi(s_r).
\end{align*}
Thus, choosing $a$ to be $\phi'(s_r)$,
(\ref{6-25-2})-(\ref{6-25-3-b}) imply
\begin{align*}
\Tr \sigman T_{n,s}
&\le e^{-nr} \\
\Tr \rhon (I-T_{n,s})
& \le e^{-n \max_{0\le s \le 1}\frac{-s r- \phi(s|\rho\|\sigma)}{1-s}}.
\end{align*}
Therefore, we obtain 
\begin{align*}
B(r|\rho\|\sigma)\ge
\max_{0\le s \le 1}\frac{-s r- \phi(s|\rho\|\sigma)}{1-s}.
\end{align*}


Let now move on to prove Lemma \ref{l1}. 
Note that the proof that we present here goes through in infinite
dimensions. The proof relies on the following quite powerful lemma.
\begin{lem}[Audenaert et al \cite{ACMMABV}]\label{lem:4}
For any two positive-semidefinite operators $A,B$
and a real number $0 \le t \le 1$,
we obtain
\begin{align*}
\Tr \{A-B \ge 0  \}B(A^t-B^t)\ge0.
\end{align*}
\end{lem}

\textit{Proof of Lemma \ref{l1}.} ---
We apply Lemma \ref{lem:4} to the case $t=s/(1-s)$, $A=X^{1-s}$ and $B=Y^{1-s}$, where $a,b$ are positive operators
and $0\le s\le 1/2$.
With $P$ the projector on the range of $(X^{1-s}-Y^{1-s})_+$, this yields
\begin{align*}
\Tr \{ X^{1-s}-Y^{1-s} \ge 0 \} Y^{1-s}(X^s-Y^s)\ge0.
\end{align*}
Subtracting both sides from $\Tr \{ X^{1-s}-Y^{1-s} \ge 0 \}(X-Y)$ 
then yields
\begin{align*}
&\Tr X^s  \{ X^{1-s}-Y^{1-s} \ge 0 \} (X^{1-s}-Y^{1-s}) \\
\le &\Tr \{ X^{1-s}-Y^{1-s} \ge 0 \}(X-Y).
\end{align*}
Since
$\{ X^{1-s}-Y^{1-s} \ge 0 \} (X^{1-s}-Y^{1-s})
\ge (X^{1-s}-Y^{1-s})$,
we have
\begin{align*}
&\Tr X - \Tr X^s Y^{1-s}
=
\Tr X^s  (X^{1-s}-Y^{1-s})\\
\le &
\Tr X^s  \{ X^{1-s}-Y^{1-s} \ge 0 \} 
(X^{1-s}-Y^{1-s}) \\
\le &
\Tr \{ X^{1-s}-Y^{1-s} \ge 0 \}(X-Y) .
\end{align*}
Using the relation
$I- \{ X^{1-s}-Y^{1-s} \ge 0 \}=
\{ X^{1-s}-Y^{1-s} < 0 \}$,
we obtain
\begin{align*}
&\Tr(I-\{ X^{1-s}-Y^{1-s} \ge 0 \}) X 
+\Tr \{ X^{1-s}-Y^{1-s} \ge 0 \}Y \\
\le&  \Tr X^s Y^{1-s}.
\end{align*}

\section{Application to Classical-Quantum Channel Coding}
As is mentioned in Hayashi-Nagaoka\cite{H-N},
the error exponent in classical-quantum channel coding
are derived from
the error exponent in simple quantum hypothesis testing.
Now, we consider the $n$-th stationary memoryless channel of 
the classical-quantum channel $x \mapsto \rho_x$.
Define the densities $R$, $S_p$ and $\sigma_p$ for a distribution $p$,
\begin{align*}
\arraycolsep=3pt
R&\defeq 
\left( \begin{array}{ccc}
p(x_1) \rho_{x_1} & &  \smash{\lower1.4ex\hbox{0}}
\\
\smash{\lower1.7ex\hbox{0}} & \ddots & \\
&& p(x_k) \rho_{x_k}
\end{array}
\right), ~\\
S_p& \defeq 
\left( \begin{array}{ccc}
p(x_1) \sigma_p
& &  \smash{\lower1.4ex\hbox{0}}
\\
\smash{\lower1.7ex\hbox{0}} & \ddots & \\
&& p(x_k) \sigma_p
\end{array}
\right) ,~
\sigma_p\defeq\sum_{x}p(x)\rho_x .
\end{align*}
In the channel coding, we usually treat the trade-off
between 
the average error probability 
${\rm P}_e(\Phi^{(n)})$ and 
the number $N$ of transmitted massages.
That is, the receiver should choose 
the recovered message among $N$ elements via the received quantum state.
This number is called the size.

Then, the inequality (44) in Hayashi-Nagaoka\cite{H-N}
mentioned that
for any distribution $p$ and any test $T^{(n)}$, 
there exists a code $\Phi^{(n)}$ with the size $N$ 
whose average error probability 
${\rm P}_e(\Phi^{(n)})$ satisfies
\begin{align}
{\rm P}_e(\Phi^{(n)})\le
2(1-\Tr R^{\otimes n}T^{(n)} )
+4 N\Tr S^{\otimes n}T^{(n)} .
\end{align}
This kind of relation between hypothesis testing and 
channel coding was
obtained by Verd\'{u} and Han \cite{Verdu-Han},
and it was researched by Han \cite{Han} more deeply\footnote{Han 
treated the exponential error rate of the channel coding 
in the original Japanese version.
However, he did not treated this topic in the English translation.}.

When $N=e^{na}$, 
applying Lemma \ref{l1} to the two cases:
$X=S_p N, Y=R$ and $Y=S_p N, X=R$,
we obtain
\begin{align}
{\rm P}_e(\Phi^{(n)})\le
4 e^{-n(sa - \varphi_p(s))}
\end{align}
for $0 \le s \le 1$, where
\begin{align}
\varphi_p(s)\defeq \log \Tr R^{1-s}S_p^s
= \log \sum_x p_x \Tr \rho_x^{1-s} \sigma_p^s.
\end{align}
This gives the exponential decreasing rate of error probability.
This upper bound improves the bound given in Hayashi-Nagaoka\cite{H-N},
which was obtained by using Ogawa-Hayashi\cite{Oga-Hay}'s 
Hoeffding bound.
Also, it can be regarded as 
the generalization Burnashev-Holevo\cite{B-H}'s result, which gives
the the exponential decreasing rate of error probability
in the pure states case.

\section{Discussions}
In this paper, we 
applied Audenaert et al\cite{ACMMABV}'s inequality to 
the Asymmetric setting of quantum hypothesis testing,
and obtained a quantum extension of Hoeffding bound
$\max_{0\le s \le 1}\frac{-s r- \phi(s|\rho\|\sigma)}{1-s}$,
which improves Ogawa-Hayashi\cite{Oga-Hay}'s bound.
We can expect that this bound is tight because 
the tightness 
of a similar bound in symmetric setting
has been showed by Nussbaum-Szko\l{}a \cite{NS}.
Further, we applied this result 
to classical-quantum channel coding and obtained a better error exponent.

\end{document}